\documentclass{PoS}

\newcommand{\be}{\begin{equation}}
\newcommand{\ee}{\end{equation}}
\newcommand{\bea}{\begin{eqnarray}}
\newcommand{\eea}{\end{eqnarray}}
\newcommand{\nn}{\nonumber}

\def\({\left(}
\def\){\right)}

\newcommand{\half}{\frac{1}{2}}

\title{Generalized Galileons for Particle Physics and Cosmology}

\ShortTitle{Generalized Galileons for Particle Physics and Cosmology}

\author{\speaker{Mark Trodden}\thanks{Parallel presentation delivered at ICHEP 2012.}\\
        Center for Particle Cosmology, Department of Physics and Astronomy,
University of Pennsylvania, Philadelphia, PA 19104\\
        E-mail: \email{trodden@physics.upenn.edu}}

\abstract{In this brief article, I summarize attempts with collaborators over the last couple of years to extend the Galileon idea in two important ways. I discuss the effective field theory construction arising from co-dimension greater than one flat branes embedded in a flat background - the multi-Galileons - and then describe symmetric covariant versions of the Galileons, more suitable for general cosmological applications. These generalized Galileons can be thought of as interesting four-dimensional field theories in their own rights, but the work described here may also make it easier to embed them into higher dimensional theories. I also briefly mention some intriguing properties, including freedom from ghosts and a non-renormalization theorem, that hint at possible applications in particle physics and cosmology.}

\FullConference{36th International Conference on High Energy Physics,\\
		July 4-11, 2012\\
		Melbourne, Australia}

\begin{document}

\section{Introduction}
The Dvali-Gabadadze-Porrati (DGP) model~\cite{Dvali:2000hr}, provides an arena in which to explore the idea of modifying gravity. Although this is fundamentally a $5$d model, it is possible to derive a $4d$ effective action by integrating out the bulk. A decoupling limit~\cite{Luty:2003vm,Nicolis:2004qq} then exists, in which the action reduces to a theory of a single scalar $\pi$, representing the position of the brane in the extra dimension, with a cubic self-interaction term $\sim (\partial\pi)^2\Box \pi$.  This term has the properties that its field equations are second order, and that it is invariant (up to a total derivative) under the transformations
\be 
\label{Galileoninvarianceold}
\pi (x) \rightarrow \pi(x) + c + b_\mu x^\mu \ ,
\ee
where $c$ and $b_{\mu}$ are constants. This symmetry has been called the {\it Galilean} symmetry.  

One may abstract this structure, to consider a $4d$ field theory with the same symmetry for the associated scalar -- the {\it Galileon}~\cite{Nicolis:2008in}.  Interestingly, there are a finite number of such {\it Galileon terms}, with fewer numbers of derivatives per field than other terms with the same symmetries.  Despite higher derivatives in the actions, the equations of motion are second order, so that no extra degrees of freedom propagate around any background.  Because of these features, there can exist regimes in which only a finite number of Galileon terms are important, and other possible terms within the effective field theory are not. This, coupled with a non-renormalization theorem for Galileons, suggests the hope of computing nonlinear effects which are exact quantum mechanically.   

In this brief summary of a parallel talk at the ICHEP 2012 meeting, I discuss extending the Galileon idea in two important directions:  the structure of multi-Galileon theories~\cite{Hinterbichler:2010xn,Deffayet:2010zh,Padilla:2010de,Padilla:2010ir,Padilla:2010tj,Zhou:2010di}, and how to covariantize the Galileon model in a way which preserves the symmetries. This leads to new $4d$ effective field theories with the same numbers of symmetries as the Galileon theories, but with different structures, with implications for cosmology and particle physics.


\section{The Galileon}

For $n\geq 1$, the Galileon Lagrangians are unique up to total derivatives and overall constants, and only the first $n$ are non-trivial in $n$-dimensions.  The tadpole term, $\pi$, is Galilean invariant, and referred to as the first-order Galileon ${\cal L}_1$, and at the first few orders, we also have 
\bea
{\cal L}_2 &=& [\pi^2] \ , \ \ \ \ {\cal L}_3 = [\pi^2][\Pi]-[\pi^3]\ , \ \ \ 
{\cal L}_4 = \half[\pi^2][\Pi]^2-[\pi^3][\Pi]+[\pi^4]-\half[\pi^2][\Pi^2], \\ \nn
{\cal L}_5&=&{1\over 6}[\pi^2][\Pi]^3-{1\over 2}[\pi^3][\Pi]^2+[\pi^4][\Pi]-[\pi^5]+{1\over 3}[\pi^2][\Pi^3] -{1\over 2}[\pi^2][\Pi][\Pi^2]+{1\over 2}[\pi^3][\Pi^2]
\ .
\eea
Here $\Pi$ is the matrix $\Pi_{\mu\nu}\equiv\partial_{\mu}\partial_\nu\pi$, $[\Pi^n]\equiv Tr(\Pi^n)$, and $[\pi^n]\equiv \partial\pi\cdot\Pi^{n-2}\cdot\partial\pi$,.  The above terms are the only ones which are non-vanishing in four dimensions. The resulting equations of motion are
\be 
\label{Galileoneom} 
{\cal E}_{n+1} \equiv -n(n+1)\eta^{\mu_1\nu_1\mu_2\nu_2\cdots\mu_n\nu_n}\left( \partial_{\mu_1}\partial_{\nu_1}\pi\partial_{\mu_2}\partial_{\nu_2}\pi\cdots\partial_{\mu_n}\partial_{\nu_n}\pi\right)=0 \ ,
\ee
where the symmetries of the $\eta$-tensor ensure that these are second order in derivatives, so no extra ghostly degrees of freedom propagate. 

All these theories are not renormalizable, i.e. they are effective theories with a cutoff. Nevertheless, there can still be regions in which they dominate, since the symmetries forbid renormalizable terms, and other terms have more derivatives.  Fascinatingly, the ${\cal L}_n$ terms are not renormalized upon loop corrections, so that their classical values can be trusted quantum-mechanically \cite{Hinterbichler:2010xn}.  


\section{Multi-Galileons and Higher co-Dimension Branes}

The Galilean symmetry can be thought of as inherited from symmetries of a probe brane floating in a flat bulk \cite{deRham:2010eu}, admitting a generalization. Consider co-dimension $N=D-d>1$, with bulk coordinates $X^A$ ranging over $D$ dimensions, and  brane coordinates  $x^\mu$ ranging over $d$ dimensions. The relevant action is invariant under Poincare transformations and world-volume reparameterization gauge symmetries 
Fixing unitary gauge, there exists a combined symmetry of the gauge fixed action that leaves the gauge fixing intact.  Its action on the remaining fields $\pi^I$ is
\be \label{multiinternalpoincare}
\delta_{P'}\pi^I=-\omega^\mu_{\ \nu}x^\nu\partial_\mu\pi^I-\epsilon^\mu\partial_\mu \pi^I+\omega^I_{\ \mu}x^\mu-\omega^\mu_{\ J}\pi^J\partial_\mu\pi^I+\epsilon^I+\omega^I_{\ J}\pi^J \ .
\ee
The first five terms are obvious generalizations of those of the single Galileon theory, while the last term is new to co-dimension greater 
than one, and corresponds to an unbroken $so(N)$ symmetry in the transverse directions.  Thus, the Poincare group $p(1,D-1)$ is broken to $p(1,d-1)\times so(N)$.  Taking the small $\pi^I$ limit, we find the extended non-relativistic internal Galilean invariance  
\be
\label{multiinternalGalilean} \delta_{P'}\pi^I=\omega^I_{\ \mu}x^\mu+\epsilon^I+\omega^I_{\ J}\pi^J \ ,
\ee
consisting of a Galilean invariance acting on each of the $\pi^I$ and an extra internal $so(N)$ rotation symmetry under which the $\pi$'s transform as a vector~\cite{Deffayet:2010zh,Padilla:2010de,Padilla:2010ir,Hinterbichler:2010xn,Zhou:2010di}.  

In $4d$, there are now only two possible actions invariant under (\ref{multiinternalGalilean});
\bea  
{\cal L}_2&=& \partial_\mu\pi^I\partial^\mu\pi_I, \\
 {\cal L}_4&=&  \partial_\mu\pi^I\partial_\nu \pi_I\left(\partial^\mu\partial_\rho\pi^J\partial^\nu\partial^\rho\pi_J-\partial^\mu\partial^\nu\pi^J\square\pi_J\right) +{1\over 2}  \partial_\mu\pi^I\partial^\mu \pi_I\left(\square\pi^J\square\pi_J-\partial_\nu\partial_\rho\pi^J\partial^\nu\partial^\rho\pi_J\right) \ . 
 \nn 
 \label{multi4thorder}
 \eea
This represents an intriguing $4d$ scalar field theory: there is a single possible interaction term, and thus a single free coupling constant  (as in, for example, Yang-Mills theory). 

In~\cite{deRham:2010eu}, it was shown how, in co-dimension one, to construct Galilean and internally relativistic invariant scalar field actions from the probe-brane prescription. To generalize this to higher co-dimensions, we need only construct an action for the embedding of a brane $X^A(x),$ which is invariant under reparametrizations and Poincare transformations.  The reparametrizations force the action to be a diffeomorphism scalar constructed out of the induced metric $g_{\mu\nu}\equiv {\partial X^A\over\partial x^\mu} {\partial X^B\over\partial x^\nu} G_{AB}(X)$, where $G_{AB}$ is the bulk metric as a function of the embedding variables $X^A$.  Poincare invariance then requires the bulk metric to be $G_{AB}(X)=\eta_{AB}$.  Fixing the gauge $X^\mu(x)=x^\mu$ then yields
\be 
g_{\mu\nu}=\eta_{\mu\nu}+\partial_\mu \pi^I\partial_\nu\pi_I \ .
\ee
The ingredients available to construct the action are the metric $g_{\mu\nu}$, the covariant derivative $\nabla_\mu$ compatible with the induced metric, the Riemann curvature tensor $R^{\rho}_{\ \sigma\mu\nu}$ corresponding to this, and the extrinsic curvature of the embedding. 
The main difference from the $5d$ case is that the extrinsic curvature now carries an extra index, $K^i_{\mu\nu}$, running over the number of co-dimensions, and associated with an orthonormal basis in the normal bundle to the brane.  In addition, the covariant derivative $\nabla_\mu$ 
has a connection that acts on the $i$ index, with an associated curvature, $R^i_{\ j\mu\nu}$.  Therefore, an action of the form
\be
\label{generalmultiaction} 
S=\left. \int d^4x\ \sqrt{-g}F\left(g_{\mu\nu},\nabla_\mu,R^{i}_{\ j\mu\nu},R^{\rho}_{\ \sigma\mu\nu},K^i_{\mu\nu}\right)\right|_{g_{\mu\nu}=\eta_{\mu\nu}+\partial_\mu \pi^I\partial_\nu\pi_I} \ ,
\ee
will have the symmetry~(\ref{multiinternalpoincare}), and its small field limit will have the Galilean invariance (\ref{multiinternalGalilean}).

The choices for the action~(\ref{generalmultiaction}) that lead to second order equations are precisely those that come from Lovelock invariants~\cite{Lovelock:1971yv} and their boundary terms~\cite{Myers:1987yn,Miskovic:2007mg}. 
The $d$-dimensional single field Galileon terms with an even number $N$ of $\pi$'s are obtained from the $(N-2)$-th Lovelock term on the brane, constructed from the brane metric. The terms with an odd number $N$ of $\pi$'s are obtained from the boundary term of the $(N-1)$-th $d+1$ dimensional bulk Lovelock term.  To extend this to higher co-dimension, we needed the corresponding higher-co-dimension boundary terms induced by the bulk Lovelock invariants. These were studied in~\cite{Charmousis:2005ey}, and the resulting $4d$ terms are surprisingly constrained, corresponding to the uniqueness of the multi-Galileon action.  

For a brane of dimension $d=4$, the prescription depends on whether the co-dimension $N$ is odd or even. If $N\neq 3$ is odd,  one obtains the dimensional continuation of the boundary terms, with the extrinsic curvature replaced by a distinguished normal component of $K^i_{\mu\nu}$.  In the special case $N=3$, things are somewhat more complicated. If $N\neq 2$ is even, the boundary term includes only a brane cosmological constant and an induced Einstein-Hilbert term.  In the special case $N=2$, the boundary terms are only 
a brane cosmological constant, and the term
\be\label{e:Neven2}
{\cal L}_{N=2} =
\sqrt{-g} \left(R[g] - (K^i)^2 + K_{\mu\nu}^i K^{\mu\nu}_{i}
 \right) \ .
\ee

The unique brane action in four dimensions for even (for simplicity) co-dimension $\geq 4$ is then
\be 
S=\int d^4 x\ \sqrt{-g}\left(-a_2+a_4 R\right) \ .
\ee
The Galileon action is obtained by substituting $g_{\mu\nu}=\eta_{\mu\nu}+\partial_\mu\pi^I\partial_\nu\pi_I$, and expanding each term to lowest non-trivial order in $\pi$.  The cosmological constant term yields an $\mathcal{O}(\pi^2)$ piece, and the Einstein-Hilbert term yields an $\mathcal{O}(\pi^4)$ piece.  Up to total derivatives, we then have 
\be 
\label{fourthorderaction} 
S=\int d^4 x\ \left[-a_2 \ \half\partial_\mu\pi^I\partial^\mu\pi_I  +a_4\ \partial_\mu\pi^I\partial_\nu\pi^J\left(\partial_\lambda\partial^\mu\pi_J\partial^\lambda\partial^\nu\pi_I-\partial^\mu\partial^\nu\pi_I\square\pi_J \right)\right] \ .
\ee
By adding a total derivative, it is straightforward to see that the $a_4$ term is proportional to the fourth order term~(\ref{multi4thorder}), and so we recover the $4d$ multi-field Galileon model. This is the unique multi-Galileon term in $4d$ and any even co-dimension. Keeping all orders in $\pi$ leads to a relativistically invariant action, a multi-field generalization of DBI with second order equations.


\section{Symmetries for Galileons on Curved Spaces}

If the Galileons are to be useful for cosmology, or form part of a more complete and dynamical picture, it is natural to consider extending them to understand their behavior on non-flat backgrounds~\cite{Goon:2011qf,Goon:2011uw,Burrage:2011bt,Goon:2011xf}.  The general context is the theory of a dynamical 3-brane moving in a fixed, but now general, (4+1)-dimensional background. The bulk has the arbitrary but fixed background metric $G_{AB}(X)$, from which we may construct the induced metric $\bar g_{\mu\nu}(x)$ and the extrinsic curvature $K_{\mu\nu}(x)$

Reparametrization-invariance of the action is guaranteed if the action is written as a diffeomorphism scalar, $F$, of $\bar g_{\mu\nu}$, $K_{\mu\nu}$, the covariant derivative $\bar\nabla_\mu$ and the curvature $\bar R^\alpha_{\ \beta\mu\nu}$ constructed from $\bar g_{\mu\nu}$.
The action has global symmetries only if the bulk metric has Killing symmetries. Given a transformation generated by a Killing vector, $K^A$, we restore our preferred gauge by making a compensating gauge transformation $\delta_{g,{\rm comp}}x^\mu=-K^\mu$.  The symmetries then combine to shift $\pi$ by
\be
\label{gaugefixsym} 
(\delta_K+\delta_{g,{\rm comp}})\pi=-K^\mu(x,\pi)\partial_\mu\pi+K^5(x,\pi) \ ,
\ee
which is a symmetry of the gauge fixed action.


For convenience, we specialize to a fluctuation that is Gaussian normal with respect to the metric $G_{AB}$, and demand that the extrinsic curvature on each of the slices be proportional to the induced metric. Under these assumptions the metric takes the form $G_{AB}dX^AdX^B=d\rho^2+f(\rho)^2g_{\mu\nu}(x)dx^\mu dx^\nu$,
where $X^5=\rho$ denotes the transverse coordinate, and $g_{\mu\nu}(x)$ is an arbitrary brane metric.  This includes all examples in which a maximally symmetric ambient space is foliated by maximally symmetric slices.  Killing symmetries which preserve the foliation will be linearly realized, whereas those that don't are realized nonlinearly.  Thus, the algebra of all Killing vectors is spontaneously broken to the subalgebra of Killing vectors preserving the foliation.

In the usual gauge the induced metric is $\bar g_{\mu\nu}=f(\pi)^2g_{\mu\nu}+\nabla_\mu\pi\nabla_\nu\pi$, and $K_{\mu\nu}$ follows. On the $4d$ brane, we can add four Lovelock and boundary terms, plus a tadpole
\bea  & {\cal L}_1=\sqrt{-g}\int^\pi d\pi' f(\pi')^4\ ,\ \ \ \ 
{\cal L}_2 = - \sqrt{-\bar g}\ ,& \ \ \ \  \nn \\
&{\cal L}_3 = \sqrt{-\bar g}K \ ,\ \ \ \ 
{\cal L}_4 =  -\sqrt{-\bar g}\bar R \ ,\ \ \ \ 
{\cal L}_5 = {3\over 2}\sqrt{-\bar g} {\cal K}_{\rm GB}& \ ,
\label{ghostfreegenterms} \eea
where the Gauss-Bonnet boundary term is
${\cal K}_{\rm GB}=-{1\over3}K^3+K_{\mu\nu}^2K-{2\over 3}K_{\mu\nu}^3-2\left(\bar R_{\mu\nu}-\half \bar R \bar g_{\mu\nu}\right)K^{\mu\nu}$. $\mathcal L_1$ is the zero derivative tadpole term which is the proper volume between any $\rho=$ constant surface and the brane position, $\pi(x)$.  While different in origin from the other terms, it too has the symmetry~(\ref{gaugefixsym}).  Each term may appear in a general Lagrangian with an arbitrary coefficient. 

There are interesting cases in which the 5d background bulk is $AdS_5$ with isometry algebra $so(4,2)$;  $dS_5$ with isometry algebra $so(5,1)$; or  $M_5$ with isometry algebra $p(4,1)$.  In addition, there are cases where the brane metric $g_{\mu\nu}$, and hence the extrinsic curvature, are maximally symmetric.  This means that the leaves of the foliation are either $4$d anti-de Sitter space $AdS_4$ with isometry algebra $so(3,2)$, 4d de-Sitter space $dS_4$ with isometry algebra $so(4,1)$, or flat 4d Minkowski space $M_4$ with isometry algebra $p(3,1)$.  In fact, there are only 6 such possible foliations, with each generating a class of theories living on an $AdS_4$, $M_4$ or $dS_4$ background and having 15 global symmetries broken to the 10 brane isometries, the same numbers as those of the original galileon.  

These simpler Lagrangians are Galileons living on curved space yet with the same number of symmetries as the full theory, whose form comes from expanding~(\ref{gaugefixsym}).  In the case of a $dS_4$ background in conformal inflationary coordinates $(u,y^i)$, the non-linear symmetries are
\be \label{dSGalileontrans}
\delta_{+}\hat\pi={1\over u}\left(u^2-y^2\right) , \ \ \
\delta_{-} \hat\pi=-{1\over u},\ \ \ 
\delta_{i} \hat\pi = {y_i\over u} \ .
\ee

A striking feature of these models is the presence of potentials with couplings determined by the symmetries~(\ref{gaugefixsym}).  In particular, $\pi$ acquires a mass of order the $dS_4$ or $AdS_4$ radius, with a value protected by the symmetries (\ref{dSGalileontrans}), so the small masses should be protected against renormalization.


\section{Summary}
Abstractions of the original Galileon idea have proven fascinating in their own rights. In this review I have discussed our work to extend them in two specific directions -- to multiple fields, and to Galileons on curved backgrounds. In both cases we have described the most general $4d$ effective theory, and how it can arise though the embedding of a suitable brane in a suitable
ambient space. The resulting symmetries are then inherited from the original bulk symmetries. 

Much of this work is quite recent, and may point the way to novel applications in cosmology, and perhaps in particle physics model building. These interesting $4d$ effective field theories have recently been shown to appear in a limit of ghost free massive gravity~\cite{deRham:2010ik,deRham:2010kj}, and an interesting open question is whether they might find a natural embedding within string theory, from whence many of the original motivations arose.

\acknowledgments
I would like to thank the organizers of ICHEP 2012 for a terrific meeting in Melbourne. I would also like to thank my collaborators - Melinda Andrews, Garrett Goon, Kurt Hinterbichler, Austin Joyce, Justin Khoury and Daniel Wesley - from whose joint work with me I have borrowed liberally in putting together this summary. This work is supported in part by the US Department of Energy, NASA ATP grant NNX11AI95G, and by the Fay R. and Eugene L. Langberg chair.


\end{document}